\documentclass{llncs}
\usepackage[ruled,vlined]{algorithm2e}
\usepackage{algorithm2e}
\usepackage{graphicx}
\usepackage{subfigure}

\newcommand{\ignore}[1]{}

\begin{document}

\title{Towards an Efficient Tile Matrix Inversion of Symmetric
  Positive Definite Matrices on Multicore Architectures}

\author{Emmanuel Agullo\inst{1} \and Henricus Bouwmeester\inst{2} \and Jack Dongarra\inst{1} \and Jakub Kurzak\inst{1} \and Julien Langou\inst{2} \and Lee Rosenberg\inst{2}}

\institute{Dpt of Electrical Engineering
    and Computer Science, University of Tennessee,\\
    1122 Volunteer Blvd, Claxton Building, Knoxville, TN 37996-3450, USA
\and
    Dpt of Mathematical and
    Statistical Sciences, University of Colorado Denver,\\
    Campus Box 170, P.O. Box 173364, Denver, Colorado 80217-3364, USA,\\
    Research was supported by the National Science Foundation grant no. NSF CCF-811520.
}

\maketitle

\begin{abstract}
  The algorithms in the current sequential numerical linear algebra libraries
(\emph{e.g.} LAPACK) do not parallelize well on multicore architectures. A new family
of algorithms, the \emph{tile algorithms}, has recently been introduced.
Previous research has shown that it is possible to write efficient and scalable
tile algorithms for performing a Cholesky factorization, a (pseudo) LU
factorization, and a QR factorization. In this extended abstract, we attack the
problem of the computation of the inverse of a symmetric positive definite
matrix. We observe that, using a dynamic task scheduler, it is relatively
painless to translate existing LAPACK code to obtain a ready-to-be-executed
tile algorithm.  However we demonstrate that non trivial compiler
techniques (array renaming, loop reversal and pipelining) need then to be
applied to further increase the parallelism of our application. We present
preliminary experimental results.

\end{abstract}
\section{Introduction}
\label{sec:introduction}

The appropriate direct method to compute the solution of a symmetric positive
definite system of linear equations consists of computing the Cholesky
factorization of that matrix and then solving the underlying triangular
systems. It is not recommended to use the inverse of the matrix in this case.
However some applications need to explicitly form the inverse of the matrix.
A canonical example is the computation of the
variance-covariance matrix in statistics. Higham~\cite[p.260,\S3]{higham} 
lists more such applications.

With their advent, multicore architectures~\cite{sutterlunch} induce
the need for algorithms and libraries that fully exploit their
capacities.  A class of such algorithms -- called tile
algorithms~\cite{Buttari2008,tileplasma} -- has been developed for one-sided dense
factorizations (Cholesky, LU and QR) and made available as part of the
Parallel Linear Algebra Software for Multicore Architectures (PLASMA)
library~\cite{plasma_users_guide}. In this paper, we extend this class
of algorithms to the case of the (symmetric positive definite) matrix
inversion. Besides constituting an important functionality for a
library such as PLASMA, the study of the matrix inversion on multicore
architectures represents a challenging algorithmic problem. Indeed,
first, contrary to standalone one-sided factorizations that have been
studied so far, the matrix inversion exhibits many
anti-dependences~\cite{compil-AK} (Write After Read). Those
anti-dependences can be a bottleneck for parallel processing, which is
critical on multicore architectures. It is thus essential to
investigate (and adapt) well known techniques used in compilation such
as using temporary copies of data to remove anti-dependences
to enhance the degree of parallelism of the matrix inversion. This
technique is known as \emph{array renaming}~\cite{compil-AK} (or
\emph{array privatization}~\cite{compil-privatization}). 
Second, \emph{loop
  reversal}~\cite{compil-AK} is to be investigated. Third, the matrix
inversion consists of three successive steps (first of which is the
Cholesky decomposition). In terms of scheduling, it thus represents an
opportunity to study the effects of \emph{pipelining}~\cite{compil-AK} those
steps on performance.

The current version of PLASMA (version 2.1) is scheduled
statically. Initially developed for the IBM Cell
processor~\cite{qr-static-scheduling-cell}, this static scheduling
relies on POSIX threads and simple synchronization mechanisms. It has
been designed to maximize data reuse and load balancing between cores,
allowing for very high performance~\cite{plasmaperf} on today's
multicore architectures. However, in the case of the matrix inversion,
the design of an ad-hoc static scheduling is a time consuming task
 and raises load balancing issues
that are much more difficult to address than for a stand-alone Cholesky
decomposition, in particular when dealing with the pipelining of
multiple steps. Furthermore, the growth of the number of
cores 
and the more 
complex memory hierarchies make executions less
deterministic.
In this paper, we rely on an experimental 
in-house dynamic scheduler~\cite{GUST}.
This scheduler is based on the idea of
expressing an algorithm through its sequential representation and
unfolding it at runtime using data hazards (Read after Write, Write
after Read, Write after Write) as constraints for parallel scheduling.
The concept is rather old and has been validated by a few successful
projects. 
We could have as well used schedulers from 
the Jade project from
Stanford University \cite{jade_1993_computer} or from the SMPSs project
from the Barcelona Supercomputer Center
\cite{cellss_2007_ibm_jrd}.

Our
discussions are illustrated with experiments conducted on a
dual-socket quad-core machine based on an Intel Xeon EMT64
processor operating at $2.26$ GHz. The theoretical
peak is equal to $9.0$ Gflop/s per core or $72.3$ Gflop/s for the
whole machine, composed of 8 cores.  
The machine is running Mac OS X 10.6.2 and is shipped with the Apple
vecLib v126.0 multithreaded BLAS~\cite{blasurl} and LAPACK vendor library, as
well as LAPACK~\cite{LAPACK_1999_guide} v3.2.1 and
ScaLAPACK~\cite{ScaLAPACK_1997_guide} v1.8.0 references.

The rest of the paper is organized as follows. In
Section~\ref{sec:inplace}, we present a new algorithm for matrix
inversion based on tile algorithms; we explain how we articulated it
with our dynamic scheduler to take advantage of multicore architectures and we compare
its performance against state-of-the-art libraries. In
Section~\ref{sec:study}, we investigate the impact on parallelism and
performance of different well known techniques used in compilation: loop
reversal, array renaming and pipelining. We conclude
and present future work directions in Section~\ref{sec:conclusion}.

\section{Tile in-place matrix inversion}
\label{sec:inplace}

Tile algorithms are a class of Linear Algebra algorithms that allow
for fine granularity parallelism and asynchronous scheduling, enabling
high performance on multicore architectures~\cite{plasmaperf,Buttari2008,tileplasma,Quintana:2009}. The
matrix of order $n$ is split into $t\times t$ square submatrices of
order $b$ ($n=b\times t$). Such a submatrix is of small granularity
(we fixed $b = 200$ in this paper) and is called a \emph{tile}. So
far, tile algorithms have been essentially used to implement one-sided
factorizations~\cite{plasmaperf,Buttari2008,tileplasma,Quintana:2009}.

\begin{figure}[htbp]
  \begin{tabular}{@{}p{.23\linewidth}p{.07\linewidth}p{.70\linewidth}@{}}
    \centering
    \subfigure[In-place (Algorithm~\ref{alg:InPlace})]{
      \label{fig:dag-inplace}
      \includegraphics[width=.23\textwidth]{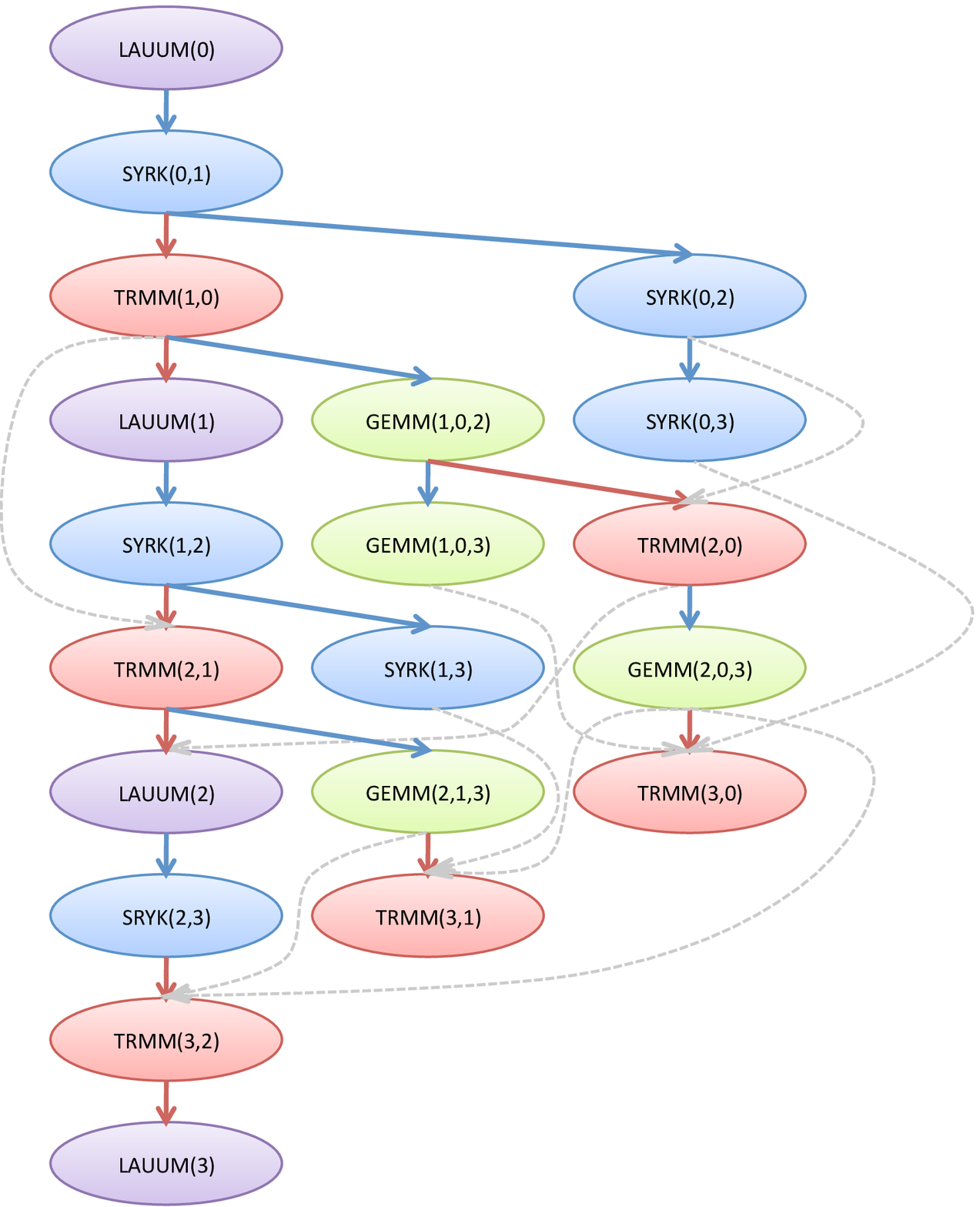}
    }%
    &
    &
    \centering
    \subfigure[Out-of-place (variant introduced in Section~\ref{sec:study})]{
      \label{fig:dag-outofplace}
      \includegraphics[width=.70\textwidth]{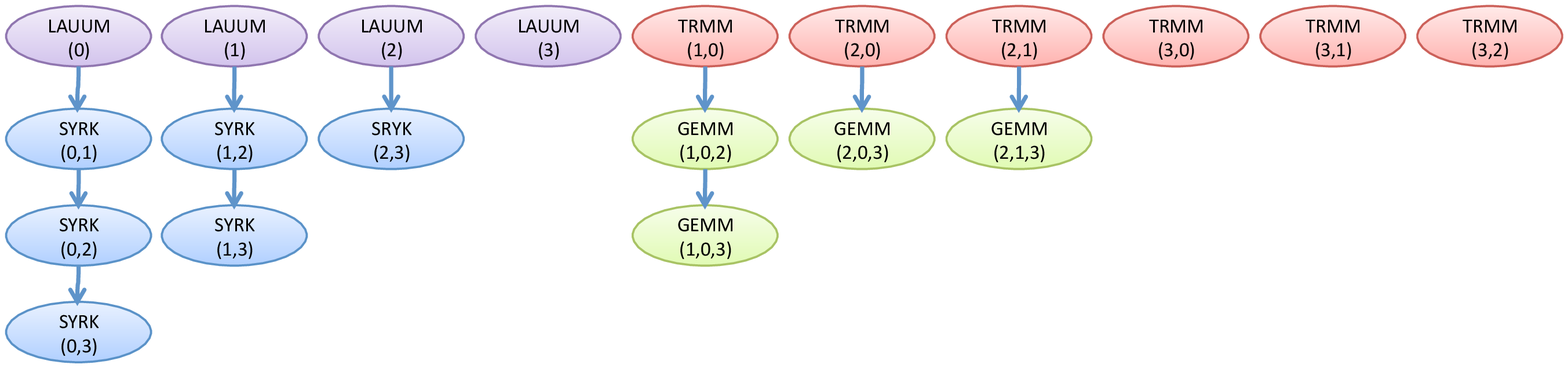}
    }%
  \end{tabular}
  \caption{DAGs of Step 3 of the Tile Cholesky Inversion ($t=4$).}
  \label{fig:dags}
\end{figure}

\begin{algorithm}
  \KwIn{$A$, Symmetric Positive Definite matrix in tile storage ($t\times t$ tiles).}
  \KwResult{$A^{-1}$, stored in-place in $A$.}
  \emph{Step~1: Tile Cholesky Factorization (compute L such that $A=LL^T$)}\;
  \For{$j=0$ \KwTo $t-1$}{
    \For{$k=0$ \KwTo $j-1$}{
      $A_{j,j} \leftarrow A_{j,j} - A_{j,k} \ast A_{j,k}^T$ (SYRK(j,k)) \;
    }
    $A_{j,j} \leftarrow CHOL(A_{j,j})$ (POTRF(j)) \;
    \For{$i=j+1$ \KwTo $t-1$}{
      \For{$k=0$ \KwTo $j-1$}{
        $A_{i,j} \leftarrow A_{i,j} - A_{i,k} \ast A_{j,k}^T$ (GEMM(i,j,k)) \;
      }
    }
    \For{$i=j+1$ \KwTo $t-1$}{
      $A_{i,j} \leftarrow A_{i,j} / A_{j,j}^T$ (TRSM(i,j)) \;
    }
  }
  \emph{Step~2: Tile Triangular Inversion of $L$ (compute $L^{-1}$)}\;
  \For{$j=t-1$ \KwTo $0$}{
    $A_{j,j} \leftarrow TRINV(A_{j,j})$ (TRTRI(j)) \;
    \For{$i=t-1$ \KwTo $j+1$}{
      $A_{i,j} \leftarrow A_{i,i} \ast A_{i,j}$ (TRMM(i,j)) \;
      \For{$k=j+1$ \KwTo $i-1$}{
        $A_{i,j} \leftarrow A_{i,j} + A_{i,k} \ast A_{k,j} $ (GEMM(i,j,k)) \;
      }
      $A_{i,j} \leftarrow - A_{i,j} \ast A_{i,i}$ (TRMM(i,j)) \;
    }
  }
  \emph{Step~3: Tile Product of Lower Triangular Matrices (compute $A^{-1}={L^{-1}}^TL^{-1}$)}\;
  \For{$i=0$ \KwTo $t-1$}{
    \For{$j=0$ \KwTo $i-1$}{
      $A_{i,j} \leftarrow A_{i,i}^T \ast A_{i,j}$ (TRMM(i,j)) \;
    }
    $A_{i,i} \leftarrow A_{i,i}^T \ast A_{i,i}$ (LAUUM(i)) \;
    \For{$j=0$ \KwTo $i-1$}{
      \For{$k=i+1$ \KwTo $t-1$}{
        $A_{i,j} \leftarrow A_{i,j} + A_{k,i}^T \ast A_{k,j}$ (GEMM(i,j,k)) \;
      }
    }
    \For{$k=i+1$ \KwTo $t-1$}{
      $A_{i,i} \leftarrow A_{i,i} + A_{k,i}^T \ast A_{k,i}$ (SYRK(i,k)) \;
    }
  }
  \caption{Tile In-place Cholesky Inversion (lower format). Matrix
    $A$ is the on-going updated matrix (in-place algorithm).}
  \label{alg:InPlace}
\end{algorithm}
Algorithm~\ref{alg:InPlace} extends this class of algorithms to the
case of the matrix inversion. As in state-of-the-art libraries
(LAPACK, ScaLAPACK), the matrix inversion is performed \emph{in-place},
\emph{i.e.}, the data structure initially containing matrix~$A$ is
directly updated as the algorithm is progressing, without using any
significant temporary extra-storage; eventually, $A^{-1}$ substitutes
$A$. Algorithm~\ref{alg:InPlace} is composed of three steps. Step~1 is
a Tile Cholesky Factorization computing the Cholesky factor $L$ (lower
triangular matrix satisfying $A=LL^T$). This step was studied
in~\cite{tileplasma}. Step~2 computes $L^{-1}$ by inverting
$L$. Step~3 finally computes the inverse matrix
$A^{-1}={L^{-1}}^TL^{-1}$. Each step is composed of multiple fine
granularity tasks (since operating on tiles). These tasks are part of
the BLAS (SYRK, GEMM, TRSM, TRMM) and LAPACK (POTRF, TRTRI, LAUUM)
standards. A more detailed description is beyond the scope of this
extended abstract and is not essential to the understanding of the
rest of the paper. Indeed, from a high level point of view, an
operation based on tile algorithms can be represented as a Directed
Acyclic Graphs (DAG)~\cite{graph} where nodes represent the fine
granularity tasks in which the operation can be decomposed and the
edges represent the dependences among them. For instance,
Figure~~\ref{fig:dag-inplace} represents the DAG of Step~3 of
Algorithm~\ref{alg:InPlace}. 

Algorithm~\ref{alg:InPlace} is based on the variants used in LAPACK 3.2.1.
Bientinesi, Gunter and van de Geijn~\cite{BientinesiGunterVanDeGeijn:08}
discuss the merit of algorithmic variations in the case of the computation
of the inverse of a symmetric positive definite matrix.  Although of definite
interest, this is not the focus of this extended abstract.

\begin{figure}[htbp]
  \begin{tabular}{@{}p{.48\linewidth}p{.04\linewidth}p{.48\linewidth}@{}}
    \centering
    \subfigure[$n = 1000$]{
      \label{fig:perf-1000}
      \includegraphics[width=.48\textwidth]{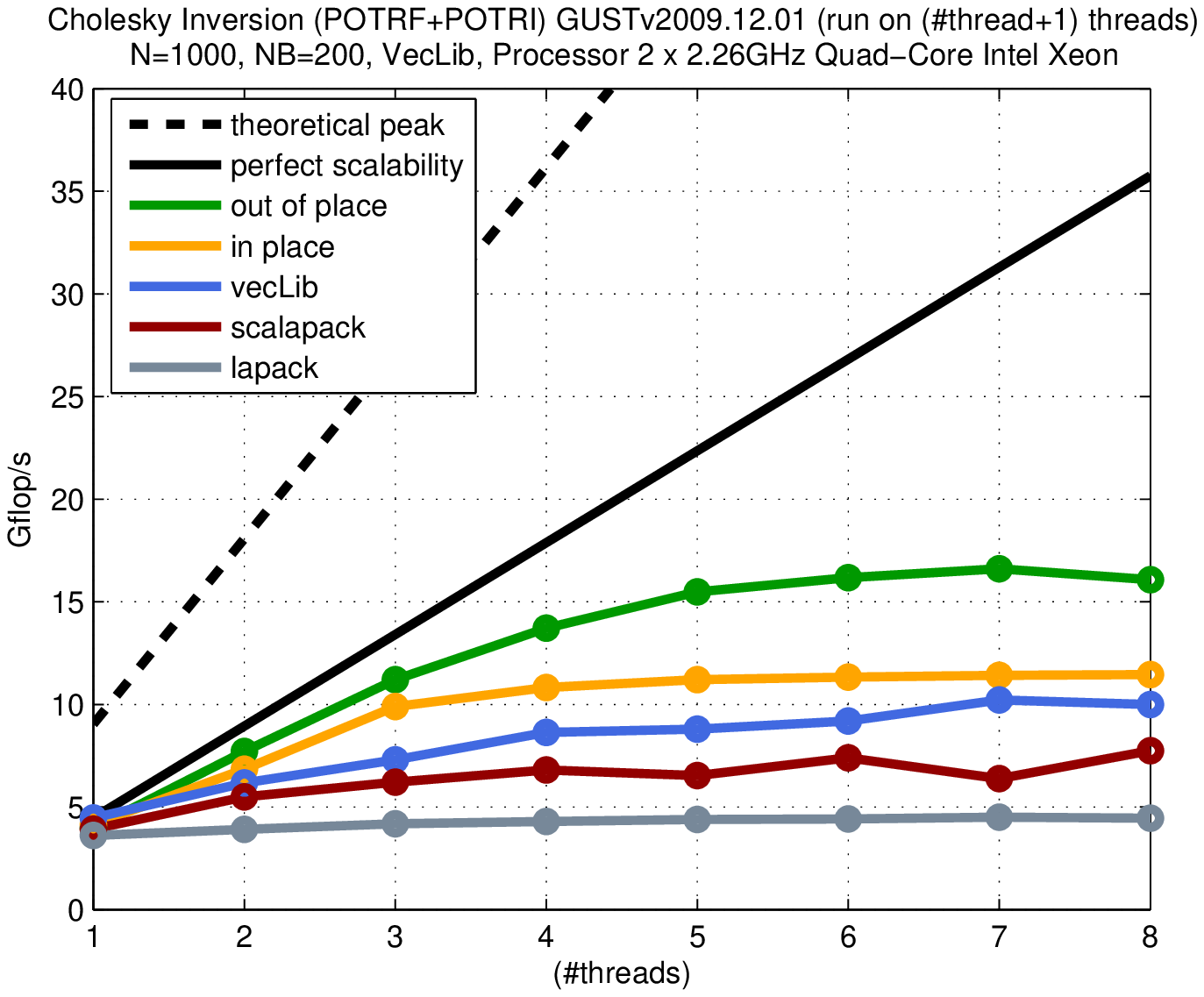}
    }%
    &
    &
    \centering
    \subfigure[$n = 4000$]{
      \label{fig:perf-4000}
      \includegraphics[width=.48\textwidth]{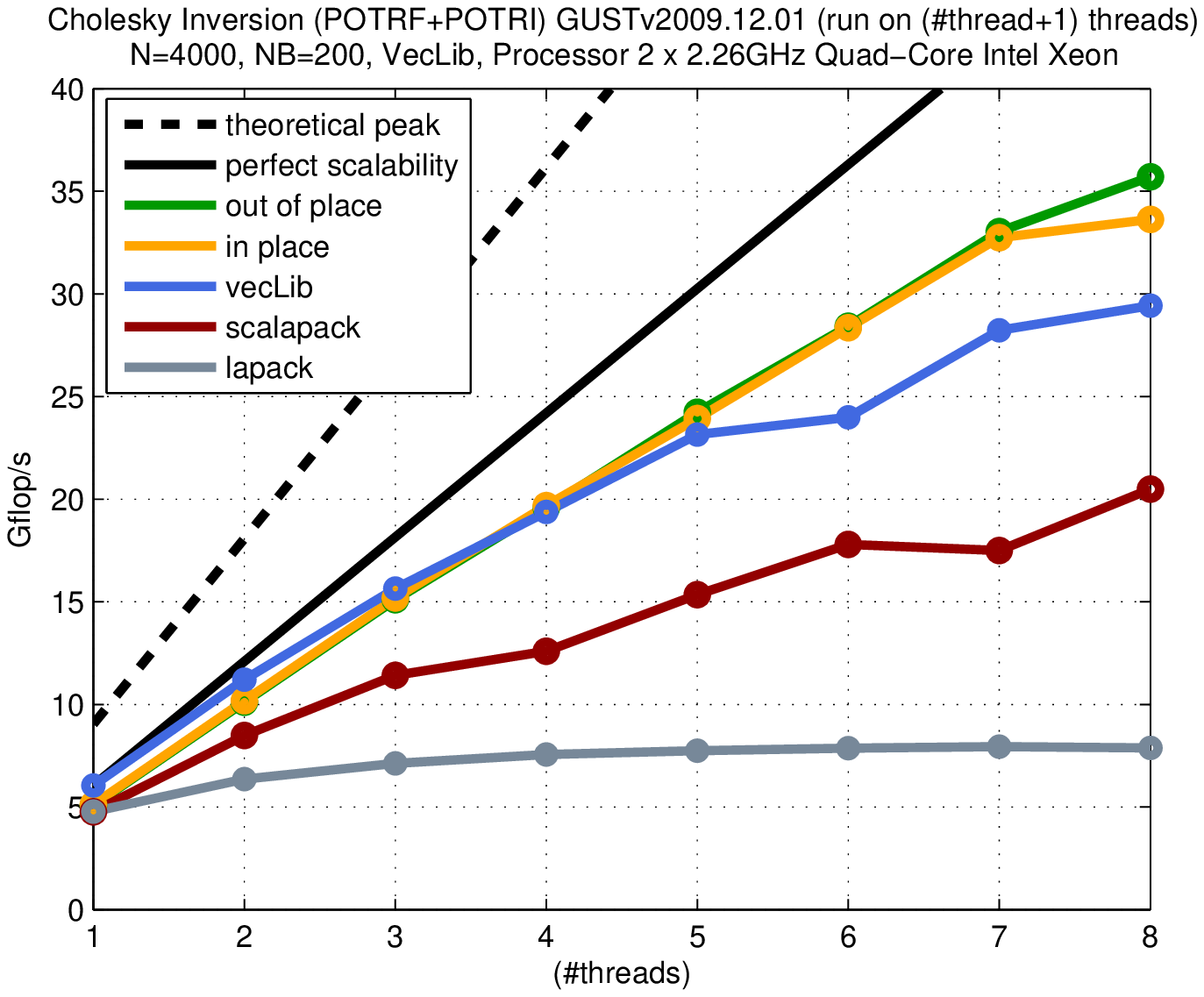}
    }%
  \end{tabular}

  \caption{Scalability of Algorithm~\ref{alg:InPlace} (in place) and
    its out-of-place variant introduced in Section~\ref{sec:study},
    using our dynamic scheduler against vecLib, ScaLAPACK and
    LAPACK libraries.}
  \label{fig:perf}
\end{figure}

We have implemented Algorithm~\ref{alg:InPlace} using our dynamic
scheduler introduced in Section~\ref{sec:introduction}.
Figure~\ref{fig:perf} shows its performance against state-of-the-art
libraries and the vendor library on the machine described in
Section~\ref{sec:introduction}. For a matrix of small size, it is
difficult to extract parallelism and have a full use of all the
cores~\cite{plasmaperf,Buttari2008,tileplasma,Quintana:2009}. We indeed observe a limited scalability
($N=1000$, Figure~\ref{fig:perf-1000}). However, tile algorithms
(Algorithm~\ref{alg:InPlace}) still benefit from a higher degree of
parallelism than blocked algorithms~\cite{plasmaperf,Buttari2008,tileplasma,Quintana:2009}. Therefore
Algorithm~\ref{alg:InPlace} (in place) consistently achieves a
significantly better performance than vecLib, ScaLAPACK and LAPACK libraries.
A larger matrix size ($N=4000$, Figure~\ref{fig:perf-4000}) allows for
a better use of parallelism. In this case, an optimized implementation
of a blocked algorithm (vecLib) competes well against tile algorithms
(in place) on few cores (left part of
Figure~\ref{fig:perf-1000}). However, only tile algorithms scale to a
larger number of cores (rightmost part of Figure~\ref{fig:perf-4000})
thanks to a higher degree of parallelism. In other words, the tile
Cholesky inversion achieves a better \emph{strong scalability} than the
blocked versions, similarly to what had been observed for the
factorization step~\cite{plasmaperf,Buttari2008,tileplasma,Quintana:2009}.

\section{Algorithmic study}
\label{sec:study}

\begin{table}[tb]
\centering
\begin{tabular}{rcccc}
\hline
           & ~~~~~ & In-place case & ~~~~~ & Out-of-place case\\%
\hline
Step 1   & & $3t-2$       & & $3t-2$          \\%
Step 2   & & $3t-3$       & & $2t-1$       \\%
Step 3   & & $3t-2$       & & $t$ \\%
\hline
\end{tabular}
\caption{Length of the critical path as a function of the number of
  tiles $t$.}
\label{table:criticalpath}
\end{table}

In the previous section, we compared the performance of the tile
Cholesky inversion against state-the-art libraries. In this section,
we focus on tile Cholesky inversion and we discuss the impact of several
variants of Algorithm~\ref{alg:InPlace} on performance.

{\bf Array renaming (removing anti-dependences).} The dependence
between SYRK(0,1) and TRMM(1,0) in the DAG of Step~3 of
Algorithm~\ref{alg:InPlace} (Figure~\ref{fig:dag-inplace}) represents
the constraint that the SYRK operation (l. 28 of
Algorithm~\ref{alg:InPlace}) needs to read $A_{k,i}=A_{1,0}$ before
TRMM (l. 22) can overwrite $A_{i,j}=A_{1,0}$. This anti-dependence
(Write After Read) can be removed thanks to a temporary copy of $A_{1,0}$.
Similarly, all the SYRK-TRMM anti-dependences, as well as TRMM-LAUMM
and GEMM-TRMM anti-dependences can be removed. 
We have designed a variant of Algorithm~\ref{alg:InPlace}
that removes all the anti-dependences thanks to the use of a large working
array (this technique is called \emph{array renaming}~\cite{compil-AK} in 
compilation~\cite{compil-AK}). 
The subsequent DAG (Figure~\ref{fig:dag-outofplace})
is split in multiple pieces (Figure~\ref{fig:dag-outofplace}), leading
to a shorter critical path (Table~\ref{table:criticalpath}).
We implemented the out-of-place algorithm, based on our dynamic scheduler
too. Figure~\ref{fig:perf-1000} shows that our dynamic scheduler exploits its higher
degree of parallelism to achieve a much higher strong scalability even
on small matrices ($N=1000$). For a larger matrix
(Figure~\ref{fig:perf-4000}), the in-place algorithm already achieved
very good scalability. Therefore, using up to $7$ cores, their performance
are similar. However, there is not enough parallelism with a
$4000\times 4000$ matrix to use efficiently all $8$ cores with the
in-place algorithm; thus the higher performance of the out-of-place
version in this case (leftmost part of Figure~\ref{fig:perf-4000}).

\ignore{TODO:8 performance graphs (4*2): 
  (step 1, step 2, step 3, pipeline) * (inplace , outofplace); 
  we can arrange in two plots: 
  - 1 plot for 6 graphs (separate steps), 
  - 1 plot for 2 graphs (pipelines)}

\begin{figure}[htbp]
  \begin{tabular}{@{}p{.48\linewidth}p{.04\linewidth}p{.48\linewidth}@{}}
    \centering
    \subfigure[$n = 1000$]{
      \label{fig:perf-loops-1000}
      \includegraphics[width=.48\textwidth]{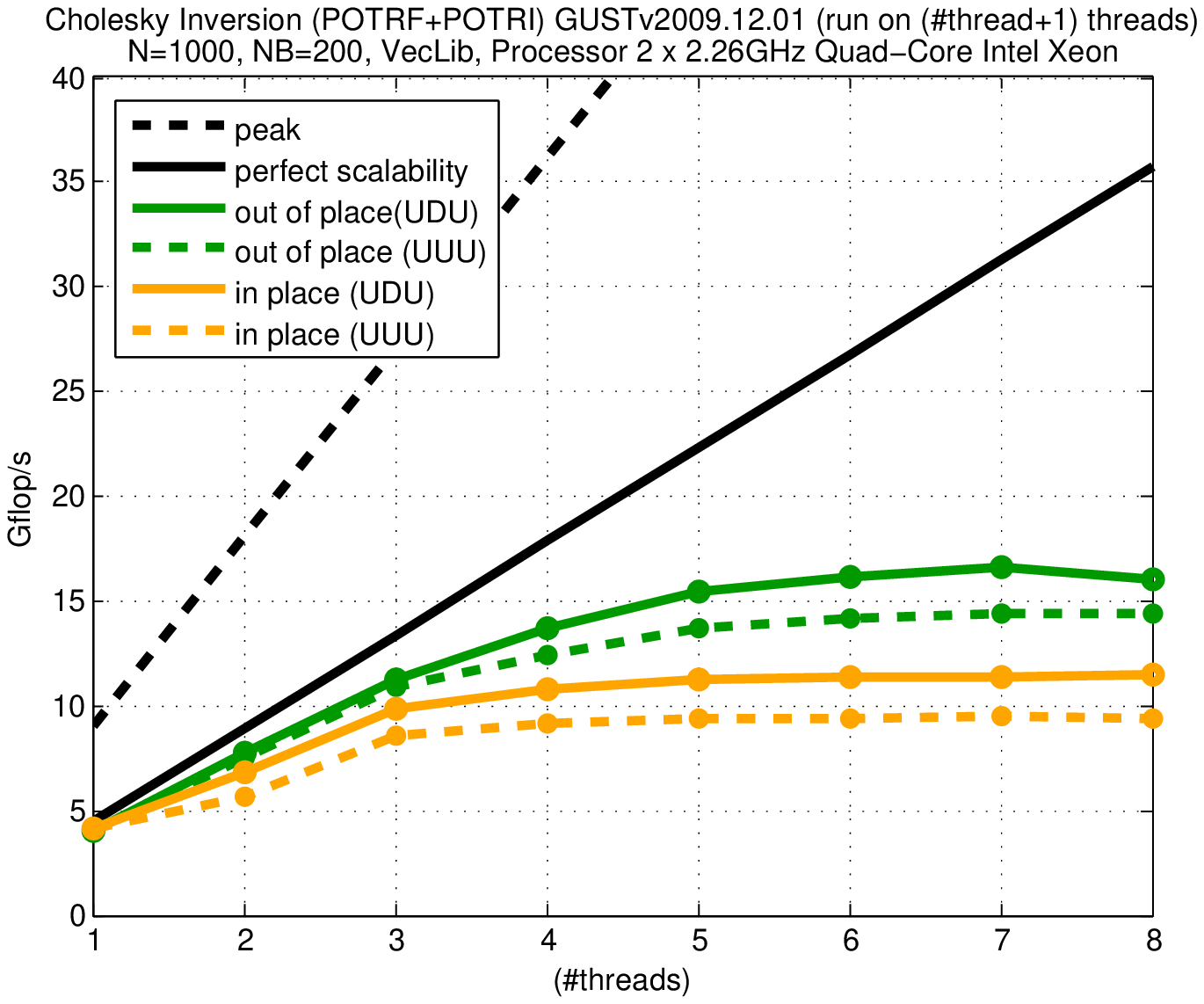}
    }%
    &
    &
    \centering
    \subfigure[$n = 4000$]{
      \label{fig:perf-loop-4000}
      \includegraphics[width=.48\textwidth]{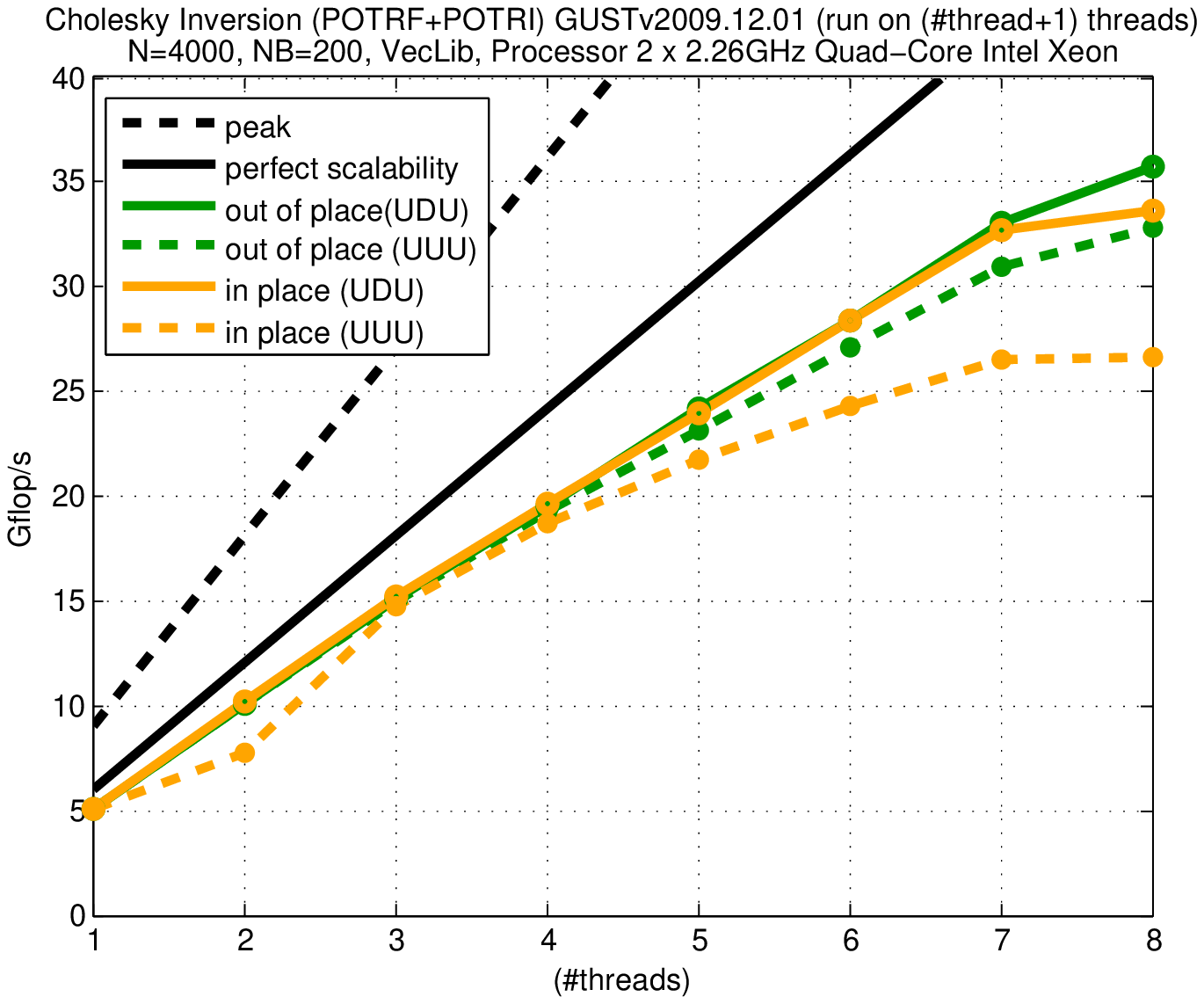}
    }\\%
  \end{tabular}
  \caption{Impact of loop reversal on performance.}
  \label{fig:perf-loops}
\end{figure}

{\bf Loop reversal (exploiting commutativity).} The most internal loop
of each step of Algorithm~\ref{alg:InPlace} (l. 88, l. 17 and l. 26)
consists in successive commutative GEMM operations. Therefore they can
be performed in any order, among which increasing order and decreasing
order of the loop index. Their ordering impacts the length of the
critical path. Algorithm~\ref{alg:InPlace} orders those three loops in
increasing (U)
and decreasing (D)
order, respectively. We had manually chosen these respective 
orders (UDU) because they minimize 
the critical path of each step (values reported in
Table~\ref{table:criticalpath}). A
naive approach would have, for example, been comprised of consistently ordering the loops
in increasing order (UUU). In this case (UUU), the critical path of TRTRI
would have been equal to $t^2-2t+3$ (in-place) or
($\frac{1}{2}t^2-\frac{1}{2}t+2$) (out-of-place) instead of $3t-3$
(in-place) or $2t-1$ (out-of-place) for (UDU). Figure~\ref{fig:perf-loops} shows
how loop reversal impacts performance.

{\bf Pipelining.} Pipelining the multiple steps of the inversion
reduces the length of its critical path. For the in-place case, the
critical path is reduced from $9t-7$ tasks ($t$ is the number of
tiles) to $9t-9$ tasks (negligible). For the out-of-place case, it is reduced from
$6t-3$ to $5t-2$ tasks. We studied the effect of pipelining on the
performance of the inversion on a $8000\times 8000$ matrix with an
artificially large tile size ($b=2000$ and $t=4$). As expected, we observed almost
no effect on performance of the in-place case (about $36.4$ seconds 
with or without pipelining).
For the out-of-place case, the elapsed time grows from
$25.1$ to $29.2$ seconds ($16~\%$ overhead) when
pipelining is prevented.

\section{Conclusion and future work}
\label{sec:conclusion}

We have proposed a new algorithm to compute the inverse of a symmetric
positive definite matrix on multicore architectures. An experimental
study has shown both an excellent scalability of our algorithm and a
significant performance improvement compared to state-of-the-art
libraries. Beyond extending the class of so-called tile algorithms,
this study brought back to the fore well known issues in the domain of
compilation. Indeed, we have shown the importance of loop reversal,
array renaming and pipelining.

The use of a dynamic scheduler allowed an out-of-the-box pipeline of
the different steps whereas loop reversal and array renaming required
a manual change to the algorithm. The future work directions consist
in enabling the scheduler to perform itself loop reversal and array
renaming. We exploited the commutativity of GEMM operations to perform
array renaming. Their associativity would furthermore allow to process them
in parallel (following a binary tree); the subsequent impact on
performance is to be studied. Array renaming requires extra-memory.
It will be interesting to address the problem of the maximization of
performance under memory constraint. This work aims to be incorporated
into PLASMA.


\begin{thebibliography}{10}

\bibitem{blasurl}
{BLAS}: {Basic} linear algebra subprograms.
\newblock \url{http://www.netlib.org/blas/}.

\bibitem{plasma_users_guide}
E.~Agullo, J.~Dongarra, B.~Hadri, J.~Kurzak, J.~Langou, J.~Langou, and
  H.~Ltaief.
\newblock {PLASMA Users' Guide}.
\newblock Technical report, ICL, UTK, 2009.

\bibitem{plasmaperf}
E.~Agullo, B.~Hadri, H.~Ltaief, and J.~Dongarrra.
\newblock Comparative study of one-sided factorizations with multiple software
  packages on multi-core hardware.
\newblock In {\em SC'09: Proceedings of the Conference on High Performance
  Computing Networking, Storage and Analysis}, pages 1--12, New York, NY, USA,
  2009. ACM.

\bibitem{compil-AK}
R.~Allen and K.~Kennedy.
\newblock {\em Optimizing Compilers for Modern Architectures: A
  Dependence-based Approach}.
\newblock Morgan Kaufmann, 2001.

\bibitem{LAPACK_1999_guide}
E.~Anderson, Z.~Bai, C.~Bischof, L.~S. Blackford, J.~W. Demmel, J.~Dongarra,
  J.~Du~Croz, A.~Greenbaum, S.~Hammarling, A.~McKenney, and D.~Sorensen.
\newblock {\em {LAPACK} Users' Guide}.
\newblock SIAM, 1992.

\bibitem{BientinesiGunterVanDeGeijn:08}
P.~Bientinesi, B.~Gunter, and R.~van~de Geijn.
\newblock Families of algorithms related to the inversion of a symmetric
  positive definite matrix.
\newblock {\em ACM Trans. Math. Softw.}, 35(1):1--22, 2008.

\bibitem{ScaLAPACK_1997_guide}
L.~S. Blackford, J.~Choi, A.~Cleary, E.~D'Azevedo, J.~Demmel, I.~Dhillon,
  J.~Dongarra, S.~Hammarling, G.~Henry, A.~Petitet, K.~Stanley, D.~Walker, and
  R.~C. Whaley.
\newblock {\em {ScaLAPACK} Users' Guide}.
\newblock SIAM, 1997.

\bibitem{Buttari2008}
A.~Buttari, J.~Langou, J.~Kurzak, and J.~Dongarra.
\newblock Parallel tiled {QR} factorization for multicore architectures.
\newblock {\em Concurrency Computat.: Pract. Exper.}, 20(13):1573--1590, 2008.

\bibitem{tileplasma}
A.~Buttari, J.~Langou, J.~Kurzak, and J.~Dongarra.
\newblock A class of parallel tiled linear algebra algorithms for multicore
  architectures.
\newblock {\em Parallel Computing}, 35(1):38--53, 2009.

\bibitem{graph}
N.~Christofides.
\newblock {\em Graph Theory: An algorithmic Approach}.
\newblock 1975.

\bibitem{compil-privatization}
R.~Eigenmann, J.~Hoeflinger, and D.~Padua.
\newblock On the automatic parallelization of the perfect
  benchmarks{\textregistered}.
\newblock {\em IEEE Trans. Parallel Distrib. Syst.}, 9(1):5--23, 1998.

\bibitem{higham}
N.~J. Higham.
\newblock {\em Accuracy and Stability of Numerical Algorithms}.
\newblock Society for Industrial and Applied Mathematics, Philadelphia, PA,
  USA, second edition, 2002.

\bibitem{GUST}
J.~Kurzak and J.~Dongarra.
\newblock Fully dynamic scheduler for numerical computing on multicore
  processors.
\newblock {\em University of Tennessee CS Tech. Report, UT-CS-09-643}, 2009.

\bibitem{qr-static-scheduling-cell}
J.~Kurzak and J.~Dongarra.
\newblock {QR factorization for the Cell Broadband Engine}.
\newblock {\em Sci. Program.}, 17(1-2):31--42, 2009.

\bibitem{cellss_2007_ibm_jrd}
J.~M. Perez, P.~Bellens, R.~M. Badia, and J.~Labarta.
\newblock {CellSs}: {Making} it easier to program the {Cell} {Broadband}
  {Engine} processor.
\newblock {\em IBM J. Res. \& Dev.}, 51(5):593--604, 2007.

\bibitem{Quintana:2009}
G.~Quintana-Ort{\'\i}, E.~S. Quintana-Ort{\'\i}, R.~A. van~de Geijn, F.~G.~Van
  Zee, and Ernie Chan.
\newblock Programming matrix algorithms-by-blocks for thread-level parallelism.
\newblock {\em {ACM} Transactions on Mathematical Software}, 36(3).

\bibitem{jade_1993_computer}
M.~C. Rinard, D.~J. Scales, and M.~S. Lam.
\newblock Jade: A high-level, machine-independent language for parallel
  programming.
\newblock {\em Computer}, 6:28--38, 1993.

\bibitem{sutterlunch}
H.~Sutter.
\newblock A fundamental turn toward concurrency in software.
\newblock {\em Dr. Dobb's Journal}, 30(3), 2005.

\end{thebibliography}
\end{document}